\documentclass[aps,twocolumn,superscriptaddress,showpacs]{revtex4}
\usepackage{epsf}
\usepackage{amsmath}

\begin{document}
\title{Thermodynamics and Kinetics of a G\={o} Protein-Like \\ Heteropolymer Model
with Two-State Folding Characteristics}
\author{Anna Kallias}
\email[E-mail: ]{Anna.Kallias@itp.uni-leipzig.de}
\altaffiliation{Present address: Volkswagen AG, Postfach 1778, D-38436 Wolfsburg, Germany}
\affiliation{Institut f\"ur Theoretische Physik and Centre for Theoretical Sciences (NTZ), 
Universit\"at Leipzig, Postfach 100\,920, D-04009 Leipzig, Germany}
\author{Michael Bachmann}
\email[E-mail: ]{Michael.Bachmann@itp.uni-leipzig.de}
\affiliation{Institut f\"ur Theoretische Physik and Centre for Theoretical Sciences (NTZ), 
Universit\"at Leipzig, Postfach 100\,920, D-04009 Leipzig, Germany}
\affiliation{Computational Biology \& Biological Physics Group, 
Department of Theoretical Physics, Lund University,
S\"olvegatan 14A, SE-223\,62 Lund, Sweden}
\author{Wolfhard Janke}
\email[E-mail: ]{Wolfhard.Janke@itp.uni-leipzig.de}
\homepage[\\ Homepage: ]{http://www.physik.uni-leipzig.de/CQT}
\affiliation{Institut f\"ur Theoretische Physik and Centre for Theoretical Sciences (NTZ), 
Universit\"at Leipzig, Postfach 100\,920, D-04009 Leipzig, Germany}
\begin{abstract}
We present results of Monte Carlo computer simulations of a coarse-grained hydrophobic-polar 
G\={o}-like heteropolymer model and discuss thermodynamic
properties and kinetics of an exemplified heteropolymer, exhibiting two-state folding
behavior. 
It turns out that general, characteristic folding features of realistic proteins with a 
single free-energy barrier
can also be observed in this simplified model, where the folding transition is primarily driven by the 
hydrophobic force.   
\end{abstract}
\maketitle
\section{Introduction}
Spontaneous protein folding is a dynamic process, which starts after the generation
of the DNA-encoded amino acid sequence in the ribosome and is in many cases finished, when the 
functional conformation, the native fold, is formed. As this process takes microseconds to
seconds, a dynamical computational analysis of an appropriate microscopic
model, which could lead to a better understanding of the conformational
transitions accompanying folding~\cite{dill0}, is extremely demanding. 
Since protein folding is a thermodynamic process at finite temperature, a certain folding
trajectory in the free-energy landscape is influenced by Brownian collisions with surrounding
solvent molecules. Therefore, it is more favorable to study the kinetics of the folding
process by averaging over an appropriate ensemble of trajectories. 

A significant problem is that the complexity of detailed semiclassical microscopic
models based on force-fields and solvent parameter sets (or explicit solvent) rules
out molecular dynamics (MD) in many cases and, therefore, Markovian Monte Carlo (MC) 
dynamics is a frequently used method for such kinetic studies. It is obvious, however, 
that the time scale provided by MC is not directly comparable with the time scale
of the folding process. It is widely believed that the folding path of a protein is
strongly correlated with contact ordering~\cite{zhou1}, i.e., the order of the successive contact 
formation between residues and, therefore, long-range correlations and memory
effects can significantly influence the kinetics.

A few years ago, experimental evidence was found that classes of proteins show 
particular simple folding characteristics, single exponential and two-state 
folding~\cite{fersht1,fersht2}.
In the two-state folding process, which is in the focus of the present study, the peptide is
either in an unfolded, denatured state or it possesses a native-like, folded structure.
In contrast to the barrier-free single-exponential folding, there exists an unstable
transition state to be passed in the two-state folding process. Due to the comparatively
simple folding characteristics, strongly simplified, effective models were established. 
Knowledge-based models of G\={o}-like type~\cite{ueda1,go1,takada1,head1} were investigated
in numerous recent 
studies~\cite{pande1,shea1,clementi1,ozkan1,cieplak1,koga1,li1,kaya1,kaya2,schonbrun1}. 
In G\={o}-like models the native fold must be known and is taken as input for the energy 
function. The energy of an actual conformation depends on its structural deviation 
from the native fold (e.g., by counting the number of already established native contacts).
By definition, the energy is minimal, if conformation and native fold are identical in all
degrees of freedom involved in the model. The simplicity of the model entails reduced
computational complexity and also MD simulations, e.g., based on Langevin dynamics~\cite{kaya1}, can
successfully be performed.

In this paper, we follow a different approach. We also study a G\={o}-like model,
but it is based on a minimalistic coarse-grained hydrophobic-polar representation of the 
heteropolymer. The basic idea behind coarse-graining is the introduction of a mesoscopic
length scale, i.e., the reduction of microscopic details, in order to classify 
heteropolymers with respect to their folding characteristics. The assumption is that,
if there is some sort of universality in folding processes, then an effective model
should allow for a general description of the qualitative folding behavior. In fact, in a recent
work we could show that two-state folding, folding through intermediates, and metastability
are inherent tertiary folding processes which can also be found in folding studies~\cite{ssbj1,ssbj2} 
of the simple hydrophobic-polar AB model~\cite{still1}. In this context, it is quite interesting
that also secondary structures are intrinsic geometries of polymer-like objects, even on a
mesoscopic scale~\cite{banavar1,banavar2,banavar3,neuhaus1}.  
\section{Model and methods}
\label{sec:mod}
In the following, we consider the hydrophobic-polar heteropolymer sequence
${\cal S}=A_3B_2\-AB_2ABA\-B_2ABABABA$, where the $A$'s indicate hydrophobic
monomers and the $B$'s polar (hydrophilic) residues. For our comparative model
study, we employ the physically motivated AB model~\cite{still1,still2} and a knowledge-based
model of G\={o} type~\cite{ueda1,go1,clementi1}, which is referred to as the G\={o}L
(G\={o}-like) model throughout the paper.

\begin{figure}
\centerline{\epsfxsize=8.8cm \epsfbox{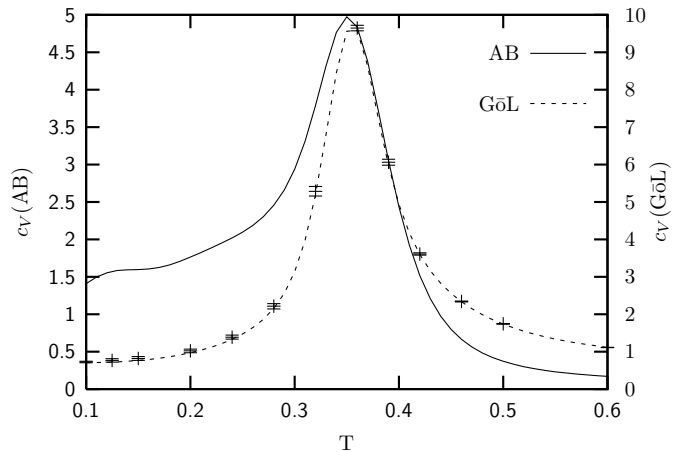}}
\caption{\label{fig:cv} 
Specific heats for the peptide with sequence ${\cal S}$ as obtained with the AB model~\cite{baj1}
and the gauged G\={o}L model as defined in Eq.~(\ref{eq:gol}).}
\end{figure}

In the AB model, the energy of a conformation $\textbf{R}=\{\textbf{r}_1,\ldots \textbf{r}_N\}$
with $N$ monomers, where $\textbf{r}_i$ denotes the spatial location of the $i$th monomer, is given by
\begin{eqnarray}
\label{eq:ab}
E_{\text{AB}}(\textbf{R}) &=& \frac{1}{4}\sum\limits_{k=1}^{N-2}(1-\cos \vartheta_k)\nonumber\\
&&\hspace{-8mm}+4\sum\limits_{i=1}^{N-2}\sum\limits_{j=i+2}^N\left(\frac{1}{r_{ij}^{12}}
-\frac{C(\sigma_i,\sigma_j)}{r_{ij}^6} \right),
\end{eqnarray}
where the first term is the bending energy and the sum runs over the $(N-2)$ bending angles 
$\vartheta_k$ ($0\le \vartheta_k\le \pi$) between monomers $k$, $k+1$, and $k+2$
of successive covalent bond vectors which have length unity. 
The Lennard-Jones type potentials depend on the types of the interacting monomers
($\sigma_i=A,B$) and on their spatial distance $r_{ij}$ and represent the influence of the 
specific AB sequence on the energy.
The long-range behavior is attractive for pairs of like monomers and repulsive for $AB$ pairs
of monomers:
\begin{equation}
\label{stillC}
C(\sigma_i,\sigma_j)=\left\{\begin{array}{cl}
+1, & \hspace{7mm} \sigma_i,\sigma_j=A,\\
+1/2, & \hspace{7mm} \sigma_i,\sigma_j=B,\\
-1/2,  & \hspace{7mm} \sigma_i\neq \sigma_j.\\
\end{array} \right.
\end{equation}
In this model, the AB heteropolymer with sequence ${\cal S}$ experiences a hydrophobic collapse 
transition which is signalized by the peak in the specific-heat curve plotted in 
Fig.~\ref{fig:cv}. The transition between random coils and native-like hydrophobic-core conformations
is a (pseudo)phase separation process and the folding transition of two-state (folded/unfolded) type.
In particular, for such systems it is known from model studies of realistic amino acid sequences
that knowledge-based G\={o} type models reveal kinetic aspects of folding and unfolding processes
reasonably well~\cite{clementi1,kaya1}. 

\begin{figure}
\centerline{\epsfxsize=4cm \epsfbox{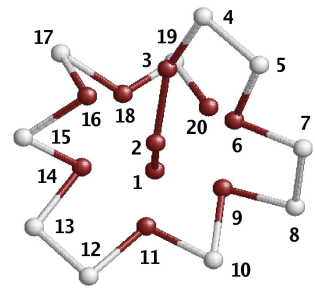}}
\caption{\label{fig:gem} 
Global-energy minimum of sequence ${\cal S}$ in the AB model with $E_{\text{AB}}\approx -19.3$.
Dark monomers are hydrophobic ($A$) and light residues polar ($B$).}
\end{figure}
For performing kinetic studies of ${\cal S}$ in the simplified hydrophobic-polar approach
as well, we use the (putative) global-energy minimum, identified in energy-landscape paving
(ELP) minimizations~\cite{hansmann1} of the AB model~\cite{baj1}, as input for the definition of a 
hydrophobic-polar G\={o}L model. The (putative) native conformation $\textbf{R}^{(0)}$ is shown in 
Fig.~\ref{fig:gem} and its energy is $E_{\text{AB}}\approx -19.3$ in the units of the AB 
model~(\ref{eq:ab}). In G\={o}L models, the ``energy'' of a given conformation is related
to its similarity with the ground state. This means, ``energy'' in the G\={o}L model plays 
rather the role of a similarity or ``order'' parameter and is, therefore, not a potential energy in the usual
physical sense (as there is no physical force associated with it). Denoting the $(N-2)$ bending angles
of the global-energy minimum conformation by $\vartheta_k^{(0)}$, its $(N-3)$ torsional angles as 
$\varphi_l^{(0)}$, and the inter-monomer distances by $r_{ij}^{(0)}$, we define the G\={o}L model
according to the representation in Ref.~\cite{clementi1} as:
\begin{eqnarray}
\label{eq:gol}
&&\hspace*{-6mm}E_{\text{G\={o}L}}(\textbf{R})/\varepsilon=K_\vartheta\sum\limits_{k=1}^{N-2}
\left(\vartheta_k-\vartheta_k^{(0)}\right)^2 \nonumber \\
&&\hspace*{-4mm}+\sum\limits_{n=1,3}\sum\limits_{l=1}^{N-3}K_\varphi^{(n)}\left\{1-
\cos\left(n\left[\varphi_l-\varphi_l^{(0)}\right]\right)\right\}   \\
&&\hspace*{-4mm}+\sum\limits_{i<j-1}^{\text{native}}
\left(5\left[\frac{r_{ij}^{(0)}}{r_{ij}} \right]^{12} 
-6\left[\frac{r_{ij}^{(0)}}{r_{ij}} \right]^{10}\right)
 +\sum\limits_{i<j-1}^{\text{nonnative}}\frac{1}{r_{ij}^{12}}.\nonumber
\end{eqnarray}
The last two sums run over all pairs of nonbonded monomers. In the case the pair $(i,j)$
forms a native contact, i.e., $r_{ij}<r_{\text{cut}}$ and $r_{ij}^{(0)}<r_{\text{cut}}$, the monomers
experience a short-range 10-12 Lennard-Jones attraction, while for nonnative contacts
an overall repulsive $1/r^{12}$ contribution is taken into account. The constants 
$K_\vartheta=20$, $K_\varphi^{(1)}=0.5$, and $K_\varphi^{(3)}=0.25$ weight the relative strengths of the 
angular energy contributions. The values were adjusted to have a reasonable coincidence
of the peak temperature of the specific heat compared to the results obtained with the AB model~\cite{baj1} 
(see Fig.~\ref{fig:cv}).
The free overall energy scale $\varepsilon$ was chosen such that 
$E_{\text{AB}}(\textbf{R}^{(0)})=E_{\text{G\={o}L}}(\textbf{R}^{(0)})=-\varepsilon\, n_{\text{tot}}$,
where $n_{\text{tot}}$ is the total number of native contacts. The definition of a native contact
requires the introduction of a cutoff radius, which we chose as $r_{\text{cut}}=1.14$. This entails
$n_{\text{tot}}=20$ for the native conformation $\textbf{R}^{(0)}$, and therefore 
$\varepsilon\approx 0.966$. The choice of the cutoff radius is not very sensitive with respect to
the results, provided it is not too small.

For the thermodynamic analyses, we performed parallel tempering simulations~\cite{huku1,huku2,geyer1} 
and for the
kinetic studies, a standard implementation of the Metropolis Monte Carlo method was used. 
Chain updates were performed employing the spherical-cap updates as described and parametrized
in Ref.~\cite{baj1}. A sweep consists of $(N-1)$ sequential bond vector update trials. 
\section{Thermodynamics}
\label{sec:thermo}
\begin{figure}
\centerline{\epsfxsize=8.8cm \epsfbox{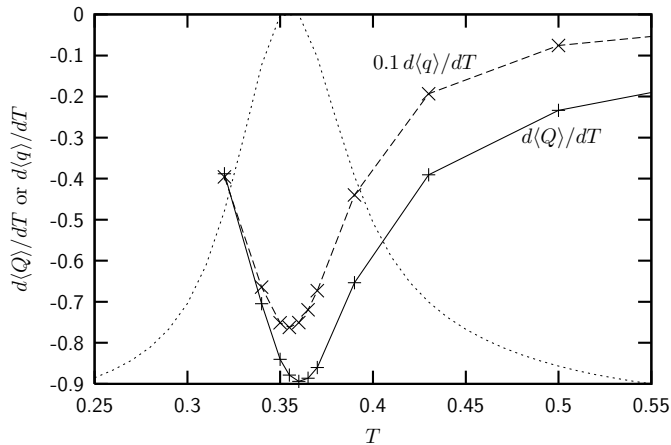}}
\caption{\label{fig:ov} 
Fluctuations of the similarity parameter $q$ and the angular overlap parameter $Q$ as
functions of the temperature. For comparison, the specific heat (dotted line)
is also plotted into this figure.}
\end{figure}
For the study of the folding transition, the introduction of an effective parameter
is useful which uniquely describes the macrostate of the ensemble of heteropolymer
conformations. A widely used measure is the contact number $q(\textbf{R})$ which is for a given
conformation $\textbf{R}$ simply defined as the fraction of the already formed native
contacts $n(\textbf{R})$ in conformation $\textbf{R}$ and the total number of native 
contacts $n_{\rm tot}$ in the final fold, i.e.,
$q(\textbf{R})=n(\textbf{R})/n_{\rm tot}$. Then, the statistical ensemble average of this quantity
$\langle q(\textbf{R})\rangle$ at a given temperature characterizes its macrostate. 
Roughly, for a two-state folder, if $\langle q(\textbf{R})\rangle > 0.5$, native-like
conformations are dominating the statistical ensemble. If less than half the
total number of contacts is formed, the heteropolymer tends to reside in the
pseudophase of denatured conformations. Note that folding transitions are not sharp phase 
transitions in the thermodynamic sense as the heteropolymer sequence is of finite length
and cannot be extended. 

Another suitable parameter is the angular overlap parameter~\cite{baj1}
which has been proven to be extremely useful in the characterization of heteropolymer
folding channels in the AB model~\cite{ssbj1,ssbj2}. 
If all virtual bond and torsion angles of two conformations, 
to be compared with each other, coincide, it is unity and between 0 and 1 otherwise. The advantage
is that it is particularly useful for classifying intermediate or metastable structures
with stable, but nonnative contacts. In Fig.~\ref{fig:ov}, the fluctuations of both 
parameters, i.e., $d\langle q\rangle/dT$ and $d\langle Q\rangle/dT$, respectively, are 
plotted as functions of the temperature for the G\={o}L model of sequence ${\cal S}$. 
We clearly see that the temperature region
of conformational activity as signalized by these two ``order'' parameters coincides
with the thermally active region indicated by the peak of the specific heat, which is also
shown for comparison. The folding temperature, i.e., the temperature of maximum activity,
is $T_f\approx 0.36$. 

\begin{figure}
\centerline{\epsfxsize=8.8cm \epsfbox{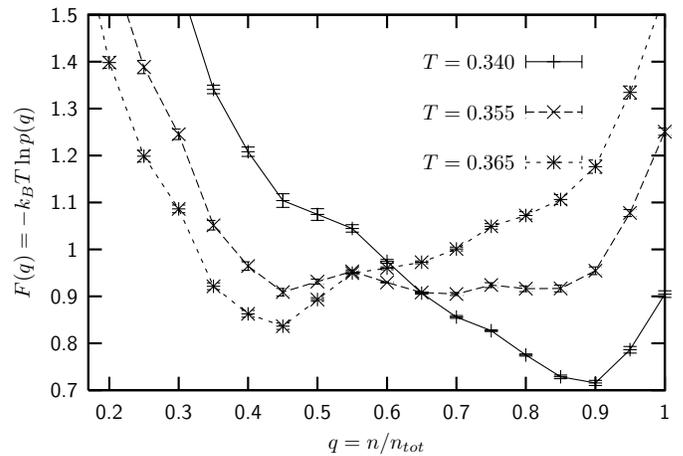}}
\caption{\label{fig:fofq} 
Free-energy landscape $F(q)$ of ${\cal S}$ for different temperatures close to the 
transition point.
}
\end{figure}
\begin{figure}
\centerline{\epsfxsize=8.0cm \epsfbox{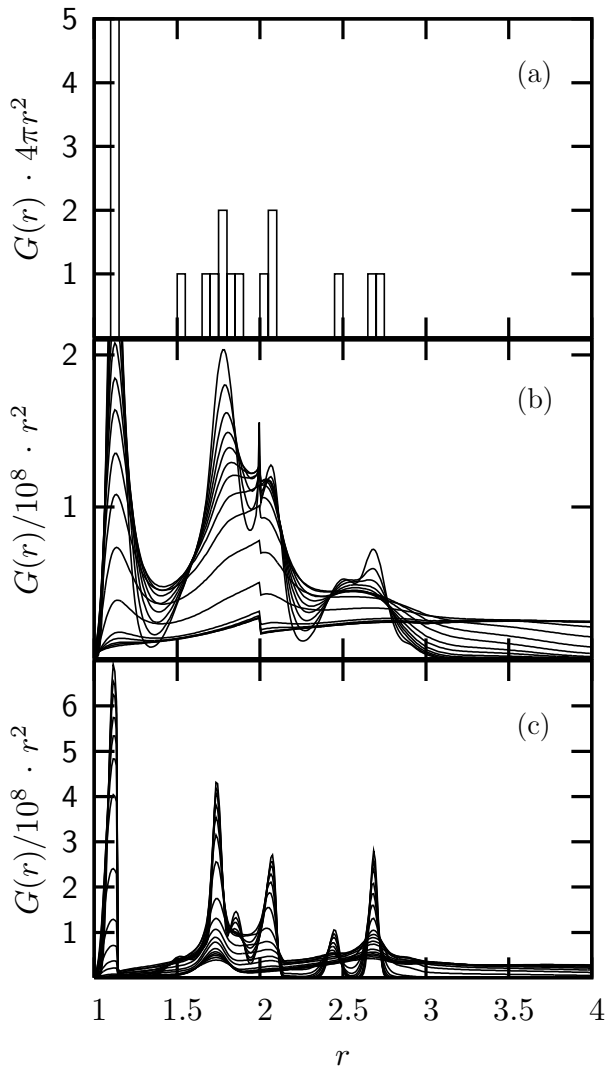}}
\caption{\label{fig:rad} 
Radial distribution functions of ${\cal S}$ (a) for the lowest-energy conformation 
(see Fig.~\ref{fig:gem}), (b) at 16 temperatures in the interval $T\in [0.1,1.5]$ (curves 
from top to bottom)
employing the AB model, and (c) at 18 temperatures in the interval $T\in [0.1,1.5]$
for the G\={o}L model.
}
\end{figure}
The classification of the heteropolymer with sequence ${\cal S}$ as two-state folder 
arises from the analysis of the free-energy landscape. We assume that $q$ is
a suitable parameter that describes the macrostate of the system adequately. Considering
this parameter as a constraint, we can formally average out the conformational
degrees of freedom and the probability for a conformation in a macrostate 
with contact parameter $q'$ reads
\begin{eqnarray}
\label{eq:pofq}
p(q')&=&\langle \delta(q'-q(\textbf{R}))\rangle \nonumber\\
&=&\frac{1}{Z}\int {\cal D}\textbf{R}\,\delta(q'-q(\textbf{R}))\,
e^{-E_\textrm{G\={o}L}(\textbf{R})/k_BT}\, , 
\end{eqnarray}
where $Z$ denotes the unrestricted partition function. The integral measure 
is simply 
${\cal D}\textbf{R}=\prod_{i=1}^N[d^3r_i]\prod_{i=1}^{N-1}[\delta(|r_{ii+1}|-1)]$. 
Expression~(\ref{eq:pofq}) can be used
to define a free energy as function of the $q$ parameter by
\begin{equation}
\label{eq:fofq}
F(q)=-k_BT \ln\,p(q).
\end{equation}
Since the value of $q$ is a qualitative measure for the macrostate the system resides in, 
the minimum of the
free energy at a given temperature $T$ is related to the dominant macrostate in the
canonical ensemble at this temperature. Actually, as can be read off from 
Fig.~\ref{fig:fofq}, the folding transition of the heteropolymer with sequence ${\cal S}$
is a phase-separation process, i.e., at the transition point close to
$T_f\approx 0.36$, the folded and the denatured pseudophase coexist and the
transition state barrier possesses a local maximum close to $q\approx 0.5$, as expected
for a typical two-state folding characteristics.

A useful measure of structure formation is the radial distribution function and its 
dependence on temperature. Due to translational invariance in the three-dimensional
space, three degrees of freedom contribute only a volume factor to the partition function.
Therefore, we utilize this by fixing the position of the first monomer, 
$\textbf{r}_1=\boldsymbol{0}$. Thus, we measure radial distances $r$ of the other monomers with 
respect to the first one and we define the radial distribution function as
\begin{equation}
\label{eq:rad}
G(r)=\frac{1}{4\pi r^2}\left\langle\delta\left(\sum_{i=3}^N[|\textbf{r}_i|-r]\right) \right\rangle\, .
\end{equation}
Note that we have excluded $i=2$ as, by definition, the virtual covalent bonds are rigid 
and, therefore, $r_{12}=1$. Actually, from our definition, 
$4\pi \int_0^\infty dr \,r^2 G(r) = N-2$. In Fig.~\ref{fig:rad}, the radial distribution
functions of ${\cal S}$ are shown for (a) the lowest-energy conformation and the ensembles
at different temperatures in the (b) AB model and (c) G\={o}L model. Although the 
global-energy minimum conformation shown in Fig.~\ref{fig:gem} does obviously not form a
regular crystal structure, the Lennard-Jones like interactions in combination with the
rigid virtual bonds induce preferable substructures, favoring, e.g., bond angles
of $60^\circ$, $90^\circ$, and $120^\circ$~\cite{ssbj2}. Therefore, the peaks of the
radial distances from the first monomer for the global energy minimum conformation 
in Fig.~\ref{fig:rad}(a) can be partly explained by these local segments. The first peak
at $r=2^{1/6}\approx 1.12$ is related to the minimum potential distance between two 
nonbonded $A$ monomers (and the reference monomer at $\textbf{r}_1=\textbf{0}$ is of type $A$). 
Also the location of
other peaks can be deduced from Fig.~\ref{fig:gem} by similar geometric arguments.

\begin{figure}
\centerline{\epsfxsize=8.8cm \epsfbox{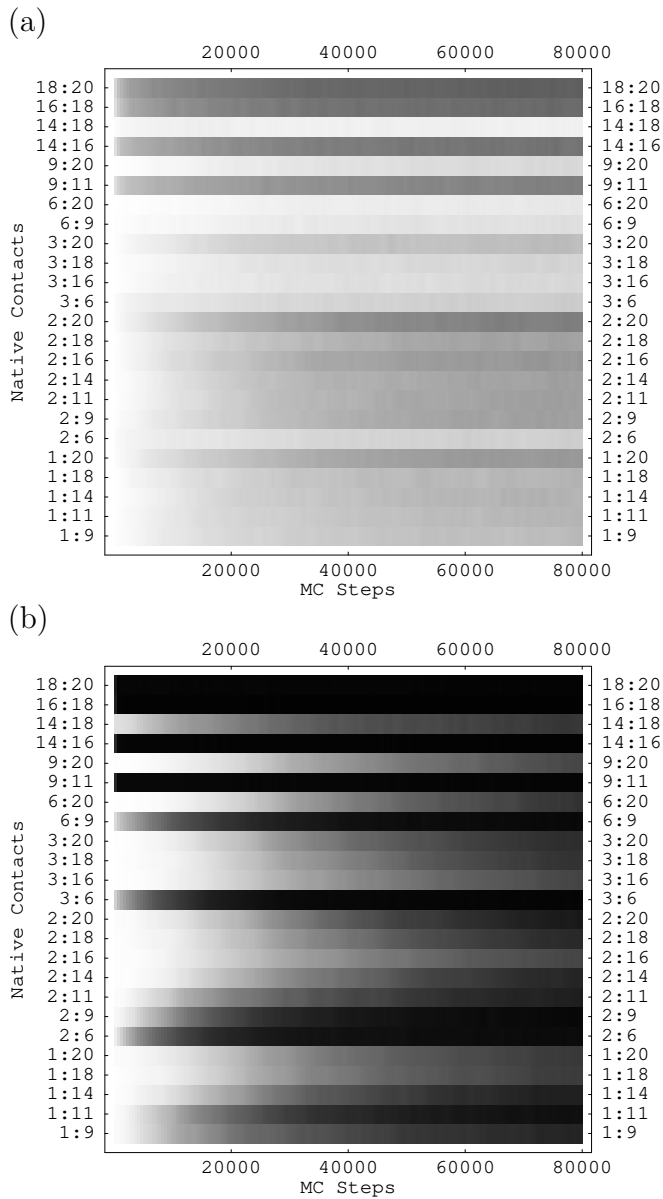}}
\caption{\label{fig:native} 
Gray-scale coded averages of the probabilities for native-contact formation of sequence ${\cal S}$ as
a function of Monte Carlo time for the (a) AB and the (b) G\={o}L model, averaged over 1\,000
folding events.  
The probability 
is calculated as a combined temporal average over 2\,000 MC steps in the ensemble of the 1\,000 
folding events. The darker the bars are, the higher is the probability that the associated contact
has formed. The labels of the pair contacts refer to the numbering in the native conformation
shown in Fig.~\ref{fig:gem}.}
\end{figure}
\begin{figure*}
\centerline{\epsfxsize=17.6cm \epsfbox{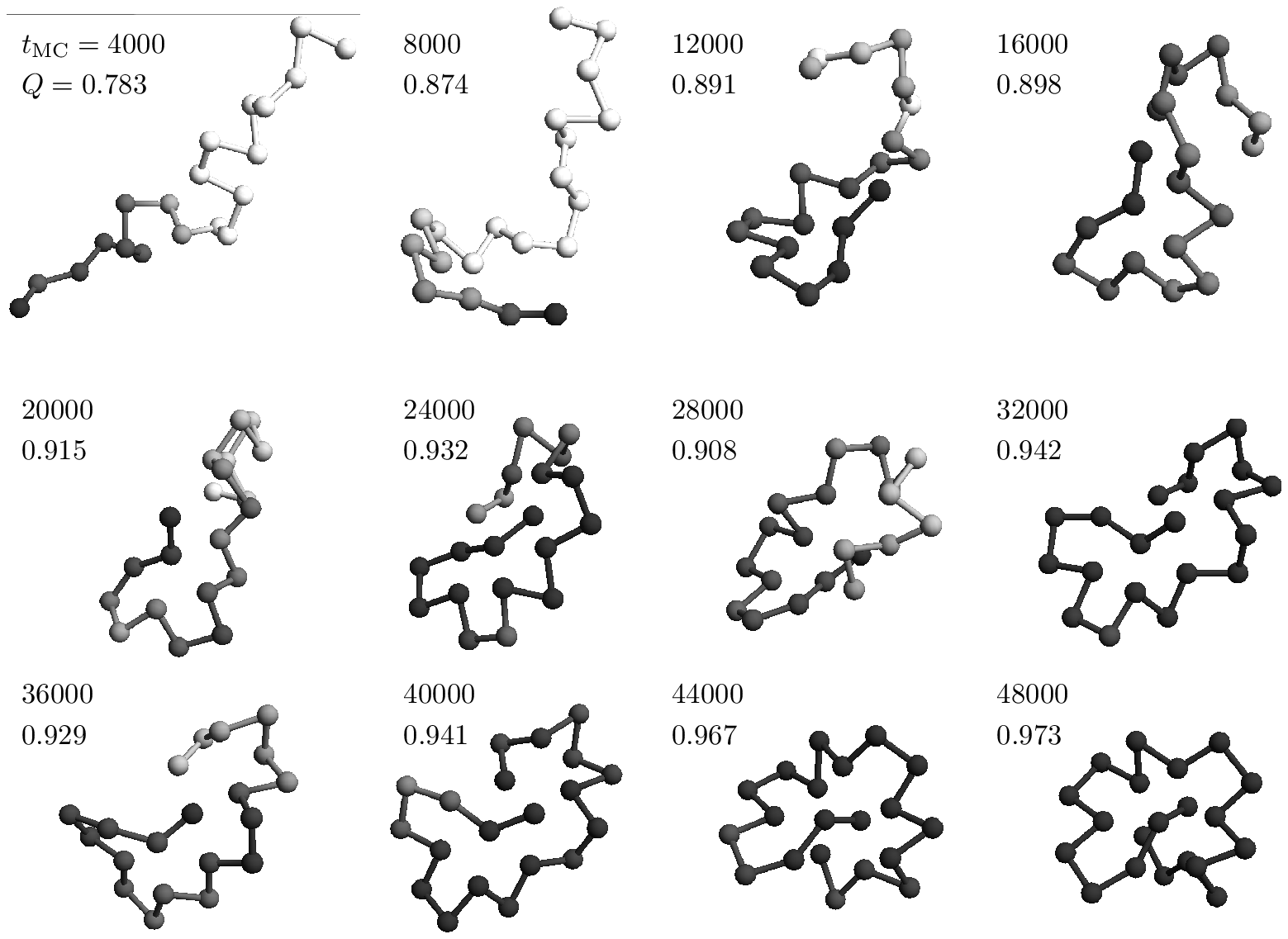}}
\caption{\label{fig:fold} 
Single folding event for the heteropolymer with sequence ${\cal S}$ in the G\={o}L model at 
$T=0.3<T_f\approx 0.36$. The
description of the grayscale code is given in the text.}
\end{figure*}
The temperature dependence of the peak evolution is shown for the AB model in 
Fig.~\ref{fig:rad}(b) and for the G\={o}L model in Fig.~\ref{fig:rad}(c). As a first
result, we see that there is nice coincidence not only in the location of the peaks,
but also in the fact that their is no indication of intermediary, weakly stable conformations.
Actually, the folding characteristics is similar in both models, at least from the 
thermodynamic point of view. At high temperatures, random-coil conformations dominate:
There is no significant structuring in the radial distribution function. Passing the
folding transition temperature, local, planar structures form first, before the 
tertiary, three-dimensional ordering towards the native conformation occurs.
\section{Kinetics}
\label{sec:kin}
The advantage of the simple G\={o}L model for two-state folding is that it enables 
likewise kinetic studies of folding {\em and} unfolding events. In fact, this 
is the main purpose of this
knowledge-based model, because kinetics studies of physically motivated models are 
typically computationally extremely demanding. This is unfortunately also the case 
for folding studies employing the AB model. A striking example is shown in 
Fig.~\ref{fig:native} where for a folding event the  gray-scale coded average probability 
of native-contact formation is plotted for the original AB model [Fig.~\ref{fig:native}(a)]
and the G\={o}L model [Fig.~\ref{fig:native}(b)]. 
The average was taken over a ``time'' window $[t_{\rm MC}-\Delta t, t_{\rm MC}+\Delta t]$ 
with $\Delta t=1\,000$ MC steps. The darker  
a bar, the higher the probability that the associated contact is formed.
As the simulation was carried out in the folding regime,
the probability of native-bond formation increases with the number of total MC
steps. It is, however, obvious that the folding is much slower using the AB model,
whereas in the G\={o}L model most of the native contacts have already been formed 
after 80\,000 MC steps.
The native conformation of the considered sequence possesses partly 
kind of zig-zag structures, or ``turns'' (see Fig.~\ref{fig:gem}). The folding of these segments is 
particularly simple and the
probability of the formation of the associated native contacts of monomers $i$ with monomers
$i+2$ or $i+3$ increases much faster than for the other contacts. But even in this case, the 
G\={o}L kinetics is unbeatable.

This example exhibits the dilemma of physics-based models in studying kinetic aspects
of structure formation by means of computer simulations. It is a notoriously difficult 
problem, because the time scale of molecular dynamics is typically too small to see folding events,
but also Markov chain Monte Carlo dynamics of physical models is typically too slow. Just for kinetic
aspects, where absolute time scales are widely irrelevant,
knowledge-based G\={o}L models are an alternative that allow for increasing the sampling efficiency, at least
for systems with simple folding characteristics (as, for example, two-state folding).
Actually, for such models an adequate 
statistical sampling of ensembles close to the transition state
can be achieved by sophisticated Monte Carlo methods where efficiency is typically
gained by simulating generalized ensembles at the expense of an artificial dynamics.  
Therefore, it is widely believed that free-energy-driven dynamics, as it is relevant for
protein folding, can reasonably be provided only by Boltzmann-Markov chains of 
conformational updates. One possibility to achieve this is the application of a
conventional Metropolis Monte Carlo method. Although the overall time scale is
left open, this method allows for comparative folding (unfolding) studies at various
temperatures. The kinetics of the folding (unfolding) transitions is thereby obtained
by averaging over sufficiently many folding (unfolding) trajectories. 
\begin{figure}
\centerline{\epsfxsize=8.8cm \epsfbox{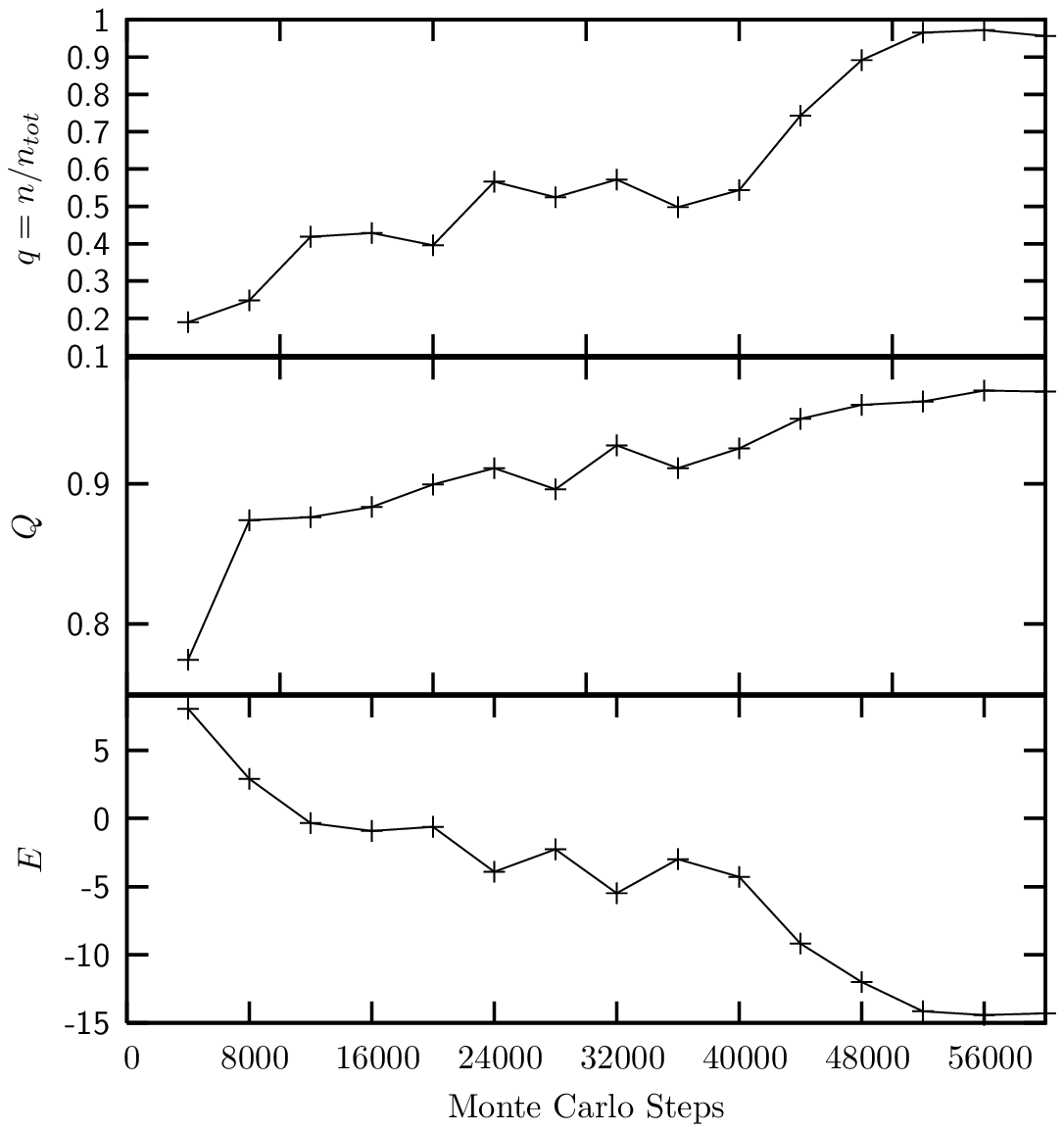}}
\caption{\label{fig:foldb} 
Temporal averages of the native-contact and the overlap similarity parameters $q$ and $Q$, respectively,
and the energy for the folding event shown in Fig.~\ref{fig:fold}. The temporal averages are calculated
every 4000 MC steps over the time interval $[t_{\rm MC}-1\,000,t_{\rm MC}+1\,000]$.}
\end{figure}

In Fig.~\ref{fig:fold}, snapshots of a single G\={o}L folding event of the heteropolymer with 
sequence ${\cal S}$ at $T=0.3$ are shown at different times $t_{\rm MC}$. The gray scale of the monomers
encodes the variance of the monomer positions,
$\sigma_{\textbf{x}_i}^2=\overline{\textbf{x}_i^2}-\overline{\textbf{x}_i}^2$,
averaged over the MC time interval $[t_{\rm MC}-1\,000,t_{\rm MC}+1\,000]$. The first monomer
is fixed and, therefore, not moving. The higher the mobility of a monomer in this time interval,
the brighter is its color. For $\sigma_{\textbf{x}_i}^2<0.1$, the monomer is rendered in black, and 
for $\sigma_{\textbf{x}_i}^2>1$ in white. The grayscales are linearly interpolated in-between these
boundaries. Although there are periods of relaxation and local unfolding, a stable intermediate
conformation is not present and the folding process is a relatively ``smooth'' process. This
is also confirmed by the more quantitative analysis of the same folding event in Fig.~\ref{fig:foldb},
where the temporal averages of the similarity parameters $q$ and $Q$, and of the energy $E$
are shown. Since the temperature lies sufficiently far below the folding temperature ($T_f\approx 0.36$),
the free-energy landscape does not exhibit substantial barriers which hinder the folding process.

\begin{figure}
\centerline{\epsfxsize=8.8cm \epsfbox{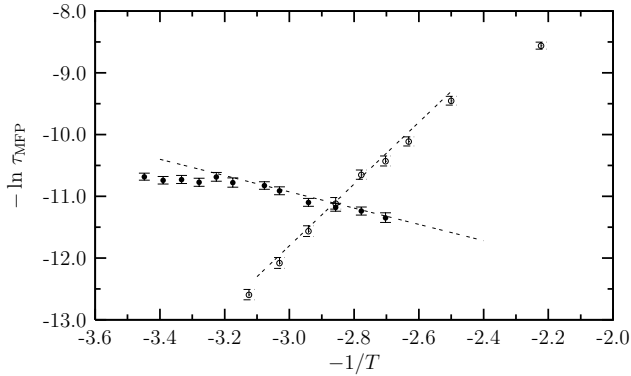}}
\caption{\label{fig:chev} Chevron plot of the mean-first passage times from
folding ($\bullet$) and unfolding ($\circ$) events at different temperatures. The hypothetic intersection point
corresponds to the transition state.}
\end{figure}

Nonetheless, the chevron plot shown in Fig.~\ref{fig:chev} exhibits a rollover which means that the
folding characteristics is not perfectly of two-state type, in which case the folding (unfolding) 
branches would be almost linear~\cite{kaya1}. In this plot, the temperature dependence of the 
mean first passage time $\tau_{\rm MFP}$ is presented. We define $\tau_{\rm MFP}$ as the average number
of MC steps that are necessary to form at least 13 native contacts in the folding 
simulations, starting from a random conformation.
In the unfolding simulations, we start from the native state and $\tau_{\rm MFP}$ is the number
of MC steps required to reach a conformation with less than 7 native contacts, i.e., 13 native contacts
are broken. 
In all simulations performed at different temperatures, $\tau_{\rm MFP}$ is averaged from the first passage
times of a few hundred respective folding and unfolding trajectories. 
Assuming a linear dependence at least in the
transition state region, $\tau_{\rm MFP}$ is directly related to exponential folding and unfolding rates
$k_{f,u}\approx 1/\tau_{\rm MFP}^{f,u}\sim \exp(-\varepsilon_{f,u}/k_BT)$, respectively where the constants 
$\varepsilon_{f,u}$ determine the kinetic folding (unfolding) propensities. The dashed lines in
Fig.~\ref{fig:chev} are tangents to the logarithmic folding and unfolding curves at the transition state.
The slopes are the folding (unfolding) propensities and have in our case values of
$\varepsilon_f\approx -1.32$ and $\varepsilon_u\approx 5.0$.

In this variant of the chevron
plot, which is similar to the presentations discussed in Refs.~\cite{kaya1,chan1}, the temperature
$T$ mimics the effect of the denaturant concentration that is in experimental studies the more generic
external control parameter. 
The hypothetic intersection
point of the folding and unfolding branches defines the transition state. The transition state 
temperature estimated from this analysis coincides very nicely with the folding temperature 
$T_f\approx 0.36$ as identified in our discussion of the thermodynamic properties of the system. 
This result also demonstrates
that the description of the folding and unfolding transitions from the kinetic point of view is
not only qualitatively, but even quantitatively consistent with the thermodynamic results from 
the canonical-ensemble analysis. 
\section{Mesoscopic heteropolymers vs.\ real proteins}
It is not the purpose of this manuscript to perform a coarse-grained analysis of the two-state folding
characteristics of a specific, real protein. Rather, we have shown that~-- although also only 
exemplified for a single sequence~-- it is actually useful to study thermodynamic and kinetic properties 
of mesoscopic peptide models without introducing atomic details. The main focus of such models is
pointed towards general features of the folding transition 
(measured in terms of ``order'' parameters being specific for the corresponding transition,
such as, e.g., the contact and overlap parameters investigated in our present study) 
that are common to a number of proteins
behaving qualitatively similarly. It is then furthermore assumed that these proteins can be grouped 
into classes of certain folding characteristics. The sequence ${\cal S}$ used in our paper is not obtained
from a one-to-one hydrophobic-polar transcription of a real amino acid sequence. We do not think
that such a mapping is particularly useful. Rather,
${\cal S}$ is considered as a representative that exhibits two-state folding
characteristics in the coarse-grained model considered here. This implies, in general, 
that the classification of peptide folding behaviors is not necessarily connected with 
detailed atomic correlations and particular contact-ordering. It is rather an intrinsic property of 
protein-like heteropolymers and can thus already be discovered employing models on mesoscopic 
scales~\cite{ssbj1,ssbj2}.

Several proteins are known to be two-state folders and their folding transitions exhibit 
the features we have also seen in our present coarse-grained model study. A famous example is 
Chymotrypsin Inhibitor 2 (CI2)~\cite{fersht1}, one of the first proteins, where two-state folding
characteristics has experimentally been identified. Clear signals of a first-order-like
folding-unfolding transition were also seen in computational G\={o} model analyses of that
peptide~\cite{clementi1,kaya1}.
It is clear, however, that a precise characterization of the transition state ensemble,
which is required for a better understanding of the folding (or unfolding) process of
a specific peptide (e.g., secondary-structure formation or disruption in CI2~\cite{li1}),
is not possible. In the models used in our study, for example, only tertiary folding aspects
based on hydrophobic-core formation are considered as being relevant. Nonetheless, as shown
in our study of sequence ${\cal S}$, macroscopic quantities or cooperativity parameters 
manifest a qualitatively similar behavior of the heteropolymer considered here, compared to real
two-state folders.
\section{Summary}
\label{sec:sum}
In this study, we have analyzed the folding thermodynamics and kinetics of a knowledge-based 
hydrophobic-polar G\={o}-like (G\={o}L) model for heteropolymers at a mesoscopic scale. For a given sequence,
the native conformation and the peak temperature of the specific heat as obtained from the 
physically motivated AB model were used to parametrize the G\={o}L model. Since the chosen 
sequence was known to exhibit two-state-like folding characteristics in the AB model, the definition of the 
corresponding G\={o}L model is appropriate for kinetic folding studies, which are 
very time consuming with the AB model itself. The reason is that the Metropolis Monte Carlo method we used
for the kinetic studies, is responsible for trapping effects and slowing down the Markovian dynamics
of the physical model. Therefore, folding events are difficult to analyze in the AB model. This is
the reason, why for statistical analyses of this model typically general-ensemble methods are 
employed~\cite{ssbj1,baj1}. Unfortunately, these methods are not suitable to study kinetic 
aspects in a fixed-temperature ensemble. 

This work focuses on the question how kinetic aspects of two-state folding behavior of realistic
proteins can also be identified in a strongly simplified mesoscopic knowledge-based model, at least
qualitatively. In Refs.~\cite{ssbj1,ssbj2}, we could show that it is possible to reveal statistical folding
properties of different folding characteristics employing the coarse-grained AB model. Here, we
have found that the corresponding 
mesoscopic G\={o}L model allows for the qualitative analysis of kinetic properties
of protein-like heteropolymers with two-state folding behavior. Thermodynamic and kinetic properties
were found to be even quantitatively consistent. 

The advantage of such simplified, coarse-grained models is 
that they enable a more global, generalized view on the physics of conformational transitions
accompanying protein folding processes. Our results show, in particular, 
that characteristic folding behaviors
are not necessarily specific to microscopic details, but also an intrinsic property of 
hydrophobic-polar heteropolymers in general.
\acknowledgments
This work is partially supported by the DFG (German Science Foundation) grant  
under contract No.\ JA 483/24-1. M.B.\ thanks the DFG and Wenner-Gren Foundation for
support by research fellowships. We acknowledge support by the DAAD-STINT Personnel Exchange
Programme. We are also grateful to the John von Neumann Institute for Computing (NIC), Forschungszentrum
J\"ulich, for the computer time grant No.\ hlz11. 
\end{document}